\documentclass[%
 aps, physrev,
 reprint,
 letterpaper,
 superscriptaddress,
 nobibnotes,
 amsmath,amssymb,
]{revtex4-2}

\usepackage{orcidlink}
\usepackage{xcolor}
\usepackage{aas_macros}
\usepackage{graphicx}
\usepackage{dcolumn}
\usepackage{bm}

\begin{document}

\title{Neutrinos as a new tool to characterise the Milky Way Centre}

\author{Paul C.~W.~Lai\,\orcidlink{0000-0003-3601-5127}}
\email{chong.lai.22@ucl.ac.uk}
\affiliation{Mullard Space Science Laboratory, University College London, Holmbury St.~Mary, Surrey RH5 6NT, United Kingdom} 

\author{Beatrice Crudele
\orcidlink{0009-0006-9431-3922}}
\email{beatrice.crudele.22@ucl.ac.uk}
\affiliation{Department of Physics and Astronomy, University College London, Gower Street, London, WC1E 6BT, United Kingdom }

\author{Matteo Agostini\,\orcidlink{0000-0003-1151-5301}}
\email{matteo.agostini@ucl.ac.uk}
\affiliation{Department of Physics and Astronomy, University College London, Gower Street, London, WC1E 6BT, United Kingdom }

\author{Hayden P.~H.~Ng\orcidlink{0009-0004-7713-2527}}
\email{ping.ng.21@ucl.ac.uk}
\affiliation{Department of Physics and Astronomy, University College London, Gower Street, London, WC1E 6BT, United Kingdom }

\author{Ellis R.~Owen\,\orcidlink{0000-0003-1052-6439}}
\email{ellis.owen@riken.jp}
\affiliation{Theoretical Astrophysics, Department of Earth and Space Science, Graduate School of Science, The University of Osaka, Toyonaka,
Osaka 560-0043, Japan}
\affiliation{Astrophysical Big Bang Laboratory (ABBL), RIKEN Pioneering Research Institute (PRI), Wak\={o}, Saitama, 351-0198 Japan}

\author{Nishta Varma}
\email{nishta.varma.23@ucl.ac.uk}
\affiliation{Department of Physics and Astronomy, University College London, Gower Street, London, WC1E 6BT, United Kingdom }

\author{Kinwah Wu\orcidlink{0000-0002-7568-8765}} 
\email{kinwah.wu@ucl.ac.uk}
\affiliation{Mullard Space Science Laboratory, University College London, Holmbury St.~Mary, Surrey RH5 6NT, United Kingdom} 

\date{\today}

\begin{abstract} 
The Central Molecular Zone (CMZ), a star-forming region rich in molecular clouds located within hundreds of parsecs from the centre of our Galaxy, converts gas into stars less efficiently than anticipated. A key challenge in refining star-formation models is the lack of precise mapping of these dense molecular hydrogen clouds, where traditional tracers often yield inconsistent results. We demonstrate how, in the near future, neutrinos will emerge as a robust mass tracer due to worldwide advancements in neutrino telescopes. Neutrinos are produced alongside gamma-rays when cosmic-rays interact with molecular clouds. The neutrino production rate is proportional to the gas density without dependence on the complex properties of a cloud. Neutrinos also have the advantage of negligible absorption and unambiguous production channels, making it a method with the lowest systematic uncertainties. In an optimistic case where most gamma-ray emission from the Galactic Centre region originates from pion decays, we expect several tens of muon neutrinos to be detected in about two decades. Neutrinos from the CMZ will provide indications on the biases of traditional mass tracers and thus indirectly enhance the accuracy of gas measurements in far galaxies from which a neutrino signal is not detectable.

\end{abstract}

\maketitle
Stars are born in vast, cold clouds of dense gas within galaxies. These clouds, composed primarily of molecular hydrogen, are known as {\it molecular clouds}~\citep{Ward-Thompson11}. The efficiency and rate of star formation in molecular clouds are governed by the physical properties of their gas, with density and total mass being the most influential factors~\citep{McKee2007ARA&A, Kennicutt2012ARA&A}. In particular, a correlation between the star-formation rate and cloud surface density has been observed~\citep{Kennicutt1998ApJ, Gao2004ApJ, Bigiel08AJ, Lada2012ApJ} and, to some extent, modelled~\citep{McKee2007ARA&A}. However, numerous star-forming clouds, both within and beyond our Galaxy, exhibit anomalous behaviour that is not accounted for by current models~\citep{Henshaw23PPVII, Kennicutt2012ARA&A}. These deviations are attributed to properties and environmental conditions of the clouds that we are currently unable to measure confidently, such as magnetisation, turbulence, and cosmic-ray heating and ionisation~\citep{Federrath2012ApJ, Padovani20SSR}. These factors not only influence star-formation properties but also introduce systematic uncertainties in measurements of the star-formation rate~\citep{Kennicutt2012ARA&A}. Thus, improving the methods used to reconstruct gas properties and characterise their systematic uncertainties is an essential step towards advancing our understanding of how galaxies convert gas into stars under different environmental and dynamical conditions.

The Central Molecular Zone (CMZ) is the inner 200-pc region at the centre of the Milky Way. This zone contains the densest clouds in our Galaxy and exhibits unique physical conditions, including high temperatures, highly turbulent motions, and strong magnetic fields~\citep{Henshaw23PPVII}. Possibly due to these factors, the star-formation rate in the CMZ is lower than the model predictions \citep{Henshaw23PPVII}.  
The proximity of the CMZ to Earth enables high-precision observations that are not possible for distant galaxies,
and the CMZ is therefore a unique laboratory to test and improve the star-formation models. It also allows the detection of weak signals that remain beyond the sensitivity limits for extragalactic sources, such as the neutrinos produced when cosmic-rays interact with the gas in molecular clouds \citep{Marinelli2017ICRC}.

In this article, we explore for the first time how neutrinos from the CMZ can offer new insights into the gas distribution by complementing more traditional probes and gamma-ray observations. These multi-messenger observations have the potential to characterise the biases and uncertainties of these traditional probes, providing a calibration method to improve the accuracy of gas distribution measurements in distant galaxies.

\begin{figure*}[tb]
\centering
\includegraphics[width=\linewidth]{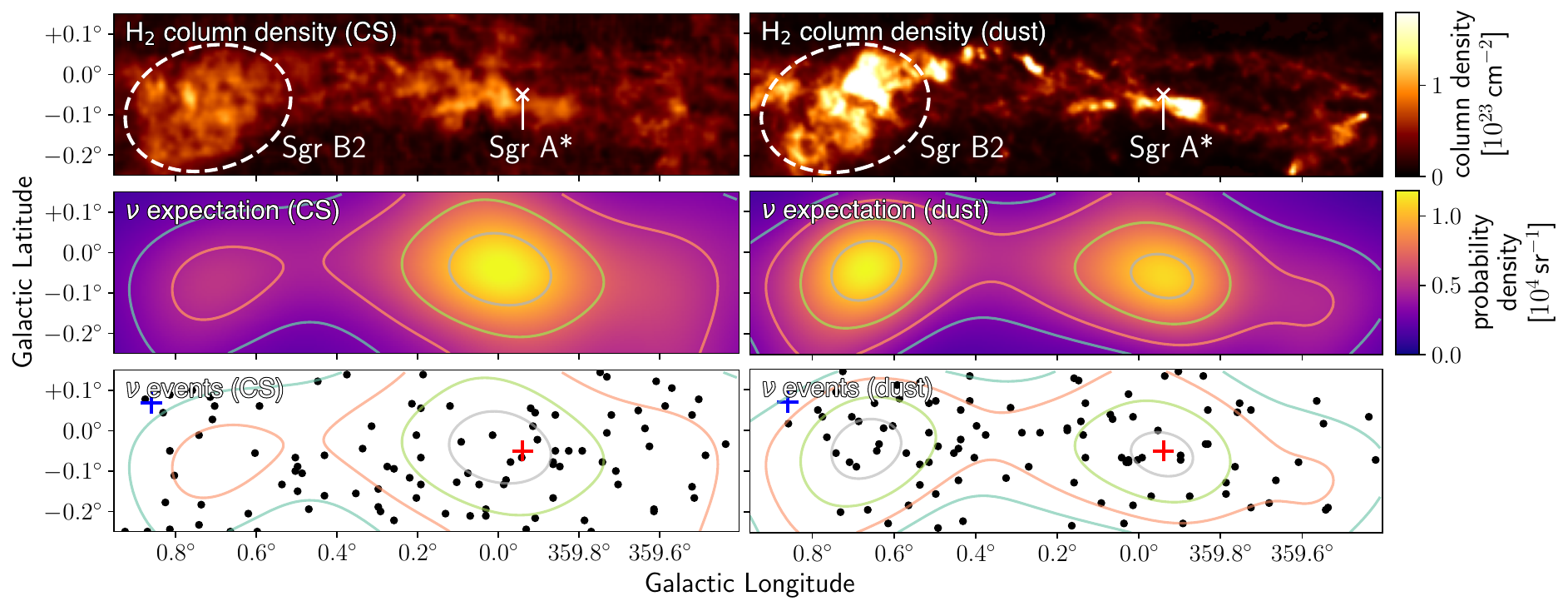}
  \caption{\emph{Top panels}: hydrogen column density in the Central Molecular Zone inferred from the CS \cite{Tsuboi99ApJS} and dust \cite{Molinari11ApJL} mass tracers, normalised to an integrated mass of $\sim10^7\,$M$_\odot$ in the displayed Galactic longitude and latitude ranges. 
  \emph{Middle panels}: probability of neutrino emission, assuming the column densities in the respective top panels, a cosmic-ray density that decreases as the inverse of the distance from Sgr A*, and an illustrative angular resolution of 0.1$^\circ$. The contours delineate iso-probability levels.
  \emph{Bottom panels}: 
  an example of how a future neutrino picture of the Galactic Centre could look like. The picture contains 100 events, which is an amount we could collect in a couple of decades by combining data from the future network of neutrino telescopes.
  The red and blue cross mark the location of HESS J1745$-$290 and G0.9+0.1, respectively.}
\label{fig:neutrino_map}
\end{figure*}

Traditional methods for determining the gas density in molecular clouds rely on indirect probes known as {\it mass tracers}. 
Unlike molecular hydrogen, which is the dominant constituent of the cloud, tracers are molecules that exist in trace abundances but emit brightly in thermal or line emission.
Observing this emission enables the gas density to be estimated through a conversion factor.
However, no single tracer is universally optimal, as each has its own characteristics and limitations. For example, carbon monoxide (CO) and dust are commonly used probes because they are abundant and emit brightly in molecular clouds, but their emission easily saturates at high densities~\cite{Kennicutt2012ARA&A, Bolatto2013ARA&A}. In contrast, tracers such as carbon monosulfide (CS) and hydrogen cyanide (HCN) emit brightly only when a cloud’s density exceeds a certain threshold, making them better suited for tracing gas above the ``critical'' density~\cite{Shirley15PASP}.

Differences between the gas measurements derived from different mass tracers are significant in the CMZ~\cite{Tsuboi99ApJS, Molinari11ApJL, Tanaka18ApJS, Tokuyama19PASJ}. 
For example, dust thermal emission indicates a very dense region in the CMZ at around $+0.7^\circ$ Galactic Longitude (known as Sgr B2), with lower densities elsewhere. On the other hand, line emission of CS suggests the gas distribution is more uniform, with the density peaking near the Galactic Centre. 
The reconstructed gas column densities (i.e. the integrated baryonic mass along the line of sight) are shown in the top panel of \figurename~\ref{fig:neutrino_map}. Reconciling different tracers has been a challenging problem, and supplementary methods are needed to characterise their uncertainties and establish a way to calibrate them. 

Further information about the gas distribution in the Galactic Centre can be extracted from gamma-rays and neutrinos generated when high-energy cosmic-rays interact with the molecular hydrogen in the cloud. Cosmic-rays injected into the CMZ, regardless of the nature or location of their sources, would diffuse due to the presence of the interstellar magnetic field and fill the CMZ. The clouds will hence act as a fixed target and collectively appear as an extended source of gamma-rays and neutrinos. The gamma-ray or neutrino surface brightness at each specific latitude and longitude of the CMZ is correlated with the column density of gas, convolved with the flux of cosmic-rays passing through it. Thus, the measured gamma-ray or neutrino flux can be directly mapped onto a gas distribution, assuming the cosmic-ray density throughout the CMZ is known.

\begin{figure}[tb]
\centering
\includegraphics[width=\linewidth]{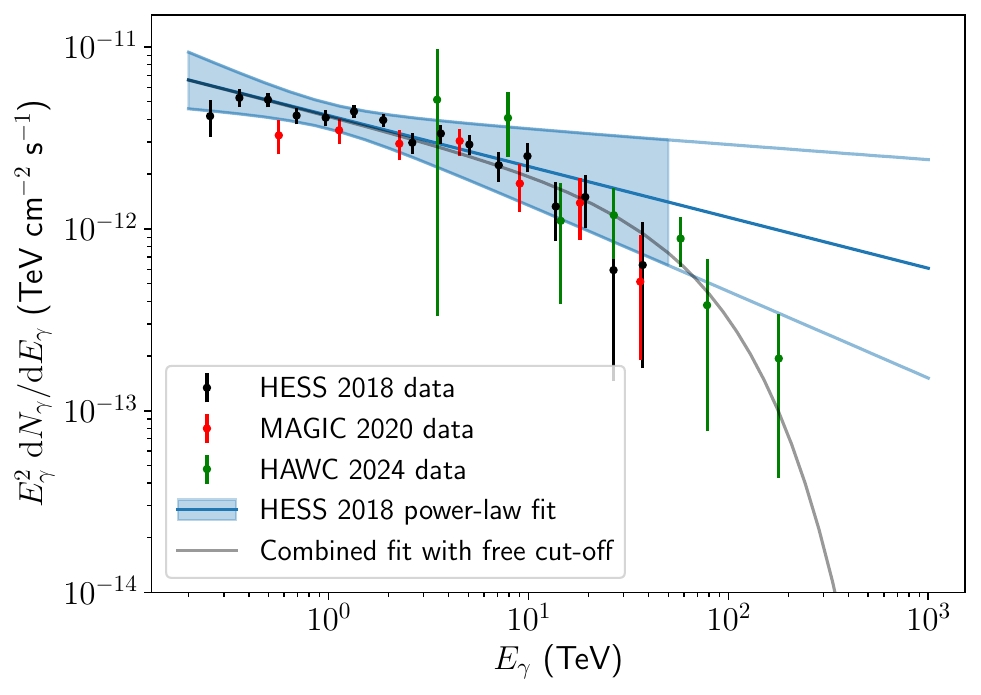} \caption{The Central Molecular Zone gamma-ray flux measured by HESS~\citep{HESS18A&A}, MAGIC~\cite{MAGIC20A&A}, and HAWC~\cite{Albert2024ApJ_HAWC}, i.e. the diffuse flux from the Galactic Centre ridge. The blue straight line shows the best fit from the HESS collaboration~\citep{HESS18A&A}, while the blue shaded areas represent our reconstructed 1$\sigma$ probability intervals, obtained by sampling the central values and uncertainties of the fit parameters. The grey line shows a maximum likelihood fit of all data available assuming a power law function with a free exponential cut-off, whose best estimate is approximately 80\,TeV.}

\label{fig:gamma_ray_obs}
\end{figure}

Over the last decade, independent observations of TeV gamma-rays from the Galactic Centre have been reported by several telescopes, including HESS~\cite{HESS16Nat, HESS18A&A}, MAGIC~\cite{MAGIC20A&A}, VERITAS~\cite{Adams2021ApJ_VERITAS}, and HAWC~\cite{Albert2024ApJ_HAWC}. 
Gamma-ray telescopes have also detected diffuse emission throughout the CMZ, often referred to Galactic Centre ridge emission within the gamma-ray community. The gamma-ray flux integrated over this spatially extended emission follows a power-law spectrum and does not display a significant energy cut-off. 
This is summarised in \figurename~\ref{fig:gamma_ray_obs}. The morphology of the gamma-ray image is compatible with expectations for the gas distribution traced by CS emission and a cosmic-ray density profile that decreases with the inverse of the distance from the Galactic Centre~\citep{HESS16Nat, HESS18A&A, MAGIC20A&A}. This cosmic-ray profile, combined with the lack of a cut-off in the gamma-ray spectrum, has been interpreted as evidence of a source capable of accelerating charged particles up to at least 1\,PeV, located within tens of parsecs of the Galactic Centre~\citep{HESS16Nat, HESS18A&A}.  

While these gamma-ray results are broadly accepted, little emphasis has been placed so far on the choice of mass tracer used to model the gas distribution. In addition, no attempt to constrain the CMZ gas distribution using gamma-ray observations has been reported in the literature.
This may be linked to intrinsic limitations that gamma-rays have in determining these quantities. Firstly, gamma-rays could be attenuated by the interstellar radiation field, which can lead to an underestimation of their flux at the source. Secondly, gamma-rays are not exclusively produced in hadronic processes, such as those resulting from cosmic-rays colliding with molecular hydrogen. They can also be produced by leptonic processes such as inverse Compton scattering. Thus, using the gamma-ray signal to map the CMZ gas distribution requires significant assumptions.

Neutrinos, on the other hand, are not absorbed, as they interact only via the weak force, and leptonic processes are not typically associated with their production at high energies. Thus, a measurement of the neutrino flux can be mapped to the gas density with more confidence than any of the methods discussed so far.

Neutrinos, however, are difficult to detect. 
Even the leading km$^3$-scale neutrino telescope, IceCube, barely detects one astrophysical neutrino from the whole sky per day \citep{IceCube-Gen22021JPhG}.
Another limitation is the field of view of IceCube, which does not include the Galactic Centre. IceCube searches requiring high angular resolution are limited to the Northern Sky, as a pure sample of ``track events'' with $<1^\circ$ angular resolution can be extracted with minimal background contamination only if the Earth acts as a filter for atmospheric muons. As a result, IceCube has only provided an initial measurement of the diffuse flux from the Galactic plane, based primarily on a sample of ``cascade events'' with an angular resolution of $\lesssim10^\circ$ above 10\,TeV~\citep{IceCube23Sci}.

Fortunately, neutrino astronomy is now at a turning point. Several neutrino telescopes capable of observing the Southern Sky with high angular resolution --- down to a tenth of a degree and below --- are being constructed or proposed worldwide. Among them, KM3NeT/ARCA 
\citep{Adrian-Martinez16JPG} and Baikal-GVD 
\citep{Baikal19arxiv} are in an advanced construction phase and are expected to begin operations within the next few years. Other proposed telescopes include P-ONE~\citep{Agostini20NatAst}, 
as well as TRIDENT~\citep{TRIDENT2023NatAs}, NEON~\citep{NEON2024arXiv} 
and HUNT~\citep{HUNT2024icrc}. 
Tentative timelines suggest that P-ONE could come online at the beginning of the next decade, followed by TRIDENT, NEON, and HUNT in the subsequent decade. This rapidly growing network of neutrino telescopes will dramatically enhance our neutrino detection capabilities in the near future and expand the field of view for high-angular-resolution neutrino observations.

To quantify the extent to which future neutrino data will help resolve discrepancies among mass tracers, we consider the scenario where neutrinos are emitted from the CMZ with the same gamma-ray flux and power-law energy spectrum as measured by HESS. 
We normalize our flux assuming that the gamma-ray flux reported in Ref.~\citep{HESS18A&A} corresponds to the integral flux over the sky window shown in \figurename~\ref{fig:neutrino_map}. We then take the best-fit parameters and uncertainties to extrapolate the neutrino energy spectra to higher energies along with its 1$\sigma$ probability interval.
The 1$\sigma$ interval is obtained by randomly sampling each fit parameter from a normal distribution centred at the best-fit value, with a standard deviation equal to the reported fit uncertainty. We then use the sampled set of parameters to build a probability distribution of the flux at each energy, with quantiles defining the desired probability intervals.
The results, shown in \figurename~\ref{fig:gamma_ray_obs}, illustrate how this construction encompasses the gamma-ray measurements performed by different telescopes across various energy ranges. 

The idea that the neutrino flux can be inferred from the gamma-ray one is motivated by the fact that all commonly considered hadronic processes result in less than 20\% difference between the gamma-ray and all-flavour neutrino flux~\cite{Koldobskiy21PRD}. This method is broadly adopted within the community. For instance, the $\pi_0$ model of Ref.~\citep{Ackermann2012ApJ_pi0} uses \emph{Fermi}-LAT's gamma-ray observations to extrapolate the integral neutrino spectrum from the entire Galactic Plane, without an energy cut-off.

As this analysis focuses on muon track events due only to muon neutrinos, we scale down the gamma-ray flux by a factor of three to account for flavour mixing by neutrino oscillations~\cite{Beacom2003PhRvD}. We then convolve the resulting fluxes with the detector response for an illustrative neutrino telescope located in the Northern Hemisphere at the site of KM3NeT, obtaining the expected reconstructed energy spectra. 
The detector response as a function of energy and neutrino arrival direction (i.e. the {\it effective area}) is taken from Ref.~\cite{Schumacher:2025qca}, which emulates the latest one from IceCube. To account for the Earth’s rotation, we perform this calculation for different times of the day, removing the periods when the Galactic Centre is outside the detector’s field of view, and then summing the results. Our calculations follow the methodology and inputs detailed in Ref.~\cite{Schumacher:2025qca}.

\begin{figure}[tb]
\centering
\includegraphics[width=\linewidth]{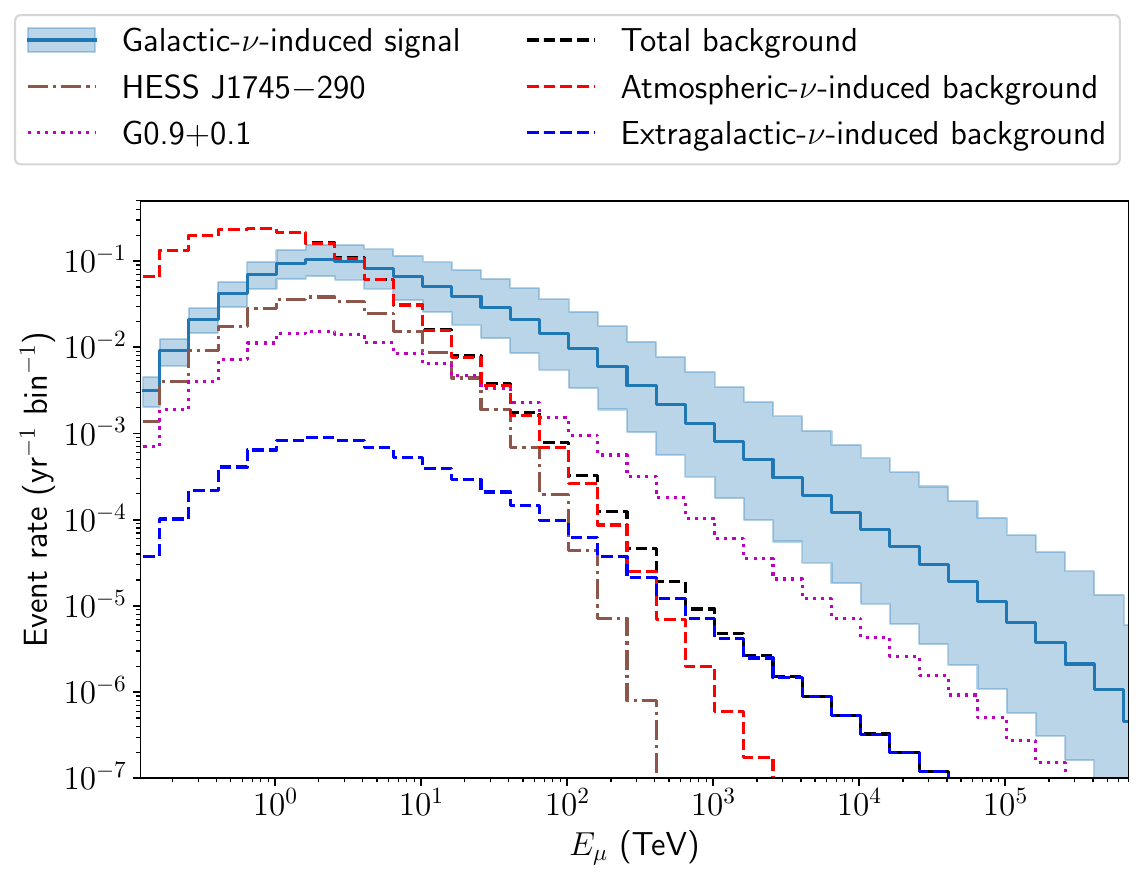}
\caption{Expected number of detected muon neutrino events from the CMZ (blue line and band), HESS J1745$-$290 (dotted green line), G0.9$+$0.1 (dotted magenta line), and background (black dashed line).
}
\label{fig:bkg_signal}
\end{figure}

Backgrounds due to diffuse astrophysical neutrinos, atmospheric neutrinos, and neutrinos from the point-like gamma-ray sources HESS J1745$-$290 and G0.9+0.1 are evaluated following the same steps described in the previous paragraph. The input atmospheric and diffuse neutrino fluxes are those measured by IceCube, while the input neutrino fluxes for HESS J1745$-$290 and G0.9+0.1 are extrapolated from the best-fit gamma-ray spectra measured by HESS \cite{Aharonian2005A&A_G09, HESS16Nat}, following the same approach used for the CMZ neutrinos. The resulting muon neutrino event rates are shown in \figurename~\ref{fig:bkg_signal}.

To keep the discussion general, we present our sensitivity studies for the future network of telescopes in terms of {\it IceCube-equivalent-years}, i.e. the effective exposure that a single detector with the efficiency of IceCube would need to achieve the combined sensitivity of the network. 
For instance, two telescopes with twice the detection efficiency of IceCube would collectively accumulate 4 IceCube-equivalent-years per year.
KM3NeT, Baikal-GVD, and P-ONE are assumed to have the same detection efficiency as IceCube, while TRIDENT, NEON, and HUNT, being larger, are assumed to correspond to 7.5, 10, and 30 IceCube-equivalent detectors, respectively.

Under these assumptions, we estimate a detection rate of $0.8$ CMZ muon neutrino events above 100\,GeV per IceCube-equivalent-year. The predicted number of events ranges from 0.5 to 1.3 when considering the $1\sigma$ fit range. 
The background contribution is expected to be 1.4 events per IceCube-equivalent-year, with
HESS J1745$-$290 and G0.9+0.1 contributing to at most 0.2 and 0.1 events, respectively.
However, the signal exceeds the background rate in specific parts of the parameter space. In particular,  as shown in \figurename~\ref{fig:bkg_signal}, CMZ neutrinos are dominant above 10\,TeV.
Additionally, as shown in \figurename~\ref{fig:neutrino_map}, the CMZ neutrinos will concentrate at regions of high gas density, while backgrounds are either uniformly distributed or peaked at the location of the gamma-ray point sources.

These estimates translate to an expected rate of about 40 muon neutrino events for 50 IceCube-equivalent-years, which is an exposure that KM3NeT, Baikal-GVD, and P-ONE collectively aim to achieve within the next 20 years. Assuming the larger telescopes come online in 2040, we anticipate collecting over 300 IceCube-equivalent-years of data within the same timeframe, resulting in several hundred muon neutrino events from the CMZ. This would mark a transformative moment for neutrino astronomy, ushering in a new era of precision measurements. An illustration of how a future neutrino image of our Galactic Centre might look is shown in the bottom panels of \figurename~\ref{fig:neutrino_map}.

While \figurename~\ref{fig:neutrino_map} demonstrates how a 2D neutrino map could provide insights into the gas distribution, the relatively small neutrino sample size raises questions about the practical significance of this information. To evaluate its impact, we conduct an illustrative statistical analysis that addresses the question: {\it What is the significance with which we will be able to discriminate between two gas distribution models?}

We frame this statistical problem as a hypothesis test, where the null hypothesis ($H_0$) assumes that the gas distribution in the CMZ follows that inferred from CS tracers, while the alternative hypothesis ($H_1$) assumes it follows the dust tracer. These two illustrative tracers were selected because they are widely used and yield the most divergent gas distributions of the CMZ, defining an envelope within which other tracers fall. 

We assess the model discrimination significance using a likelihood-ratio test statistic:
\begin{equation}
    \Lambda = \frac
    {{\rm max}_{\bm{\theta}} \,\,\mathcal{L}
    (\bm{ x}|\bm{{\theta}}, H_0)}
    {{\rm max}_{\bm{\theta}} \,\,\mathcal{L}
    (\bm{ x}|\bm{{\theta}}, H_1)} \ ,
\end{equation}
where $\bm{ x} = \{\bm{x}_1,\ldots,\bm{x}_N\}$ is a vector of data associated to the $N$ events in the data set. For each event, the data come as a vector of energy, longitude and latitude values, $\bm{x}_i=\{E_i,\,l_i,\,b_i\}$.
$\bm{\theta} = \{n_s,\,n_b,\,n_{\rm p1},\,n_{\rm p2}\}$ represents the fit parameters, 
    i.e. the numbers of events due to the CMZ, background, 
    HESS 1745$-$290, and G0.9+0.1, respectively.
The likelihood is defined as the product of the probability that each event seen comes from either model in question: 
\begin{equation}
    \mathcal{L} = \prod_{i=1}^N \frac
    {n_s\, f_s(\bm{ x}_i) +
    n_b\, f_b(\bm{ x}_i) +
    n_{\rm p1}\,f_{\rm p1}(\bm{ x}_i) +
    n_{\rm p2}\,f_{\rm p2}(\bm{ x}_i)} 
    {N}
\end{equation}
where $f_s$ refers to the probability distribution function for the signal, which differs for the null and alternative hypothesis.
$f_b$, $f_{\rm p1}$, and $f_{\rm p2}$
are the probability distribution function for the background and the two point sources.
The four numbers of events used to scale the probability distributions are treated as nuisance parameters.

The probability distributions used in the likelihood have four dimensions: energy, angular resolution, latitude and longitude. The angular resolution as a function of detected muon energy is taken from Ref.~\cite{KM3NeT2024EPJC}, and used to smear the latitude-longitude maps with a Gaussian kernel. Since we assume a univocal relationship between energy and angular resolution, the probability distribution becomes effectively three dimensional. 
The middle panels of \figurename~\ref{fig:neutrino_map} show the distribution with an angular resolution of $0.1^\circ$ as an illustration. For atmospheric and diffuse astrophysical neutrinos we assume a flat distribution in latitude and longitude. For HESS J1745$-$290 and G0.9+0.1, point-like emission is assumed, centred at the source locations marked in \figurename~\ref{fig:neutrino_map}. Lastly, the energy probability distributions are those shown in \figurename~\ref{fig:bkg_signal}.

Given the small sample size, we employ a parametric bootstrapping technique to generate the probability distribution of the test statistic under each hypothesis. To do so, we generate datasets under both hypotheses by sampling signal and background events from the same four-dimensional probability distributions used in the likelihood, and then compute the test on each dataset. Finally, we extract the model-discrimination significance by identifying the quantile of the test statistic’s probability distribution under the alternative hypothesis that corresponds to the median of the null hypothesis distribution. We perform this bootstrapping procedure for a range of IceCube-equivalent-year values to interpolate the evolution of discrimination significance over time.

\begin{figure}
\centering
\includegraphics[width=\linewidth]{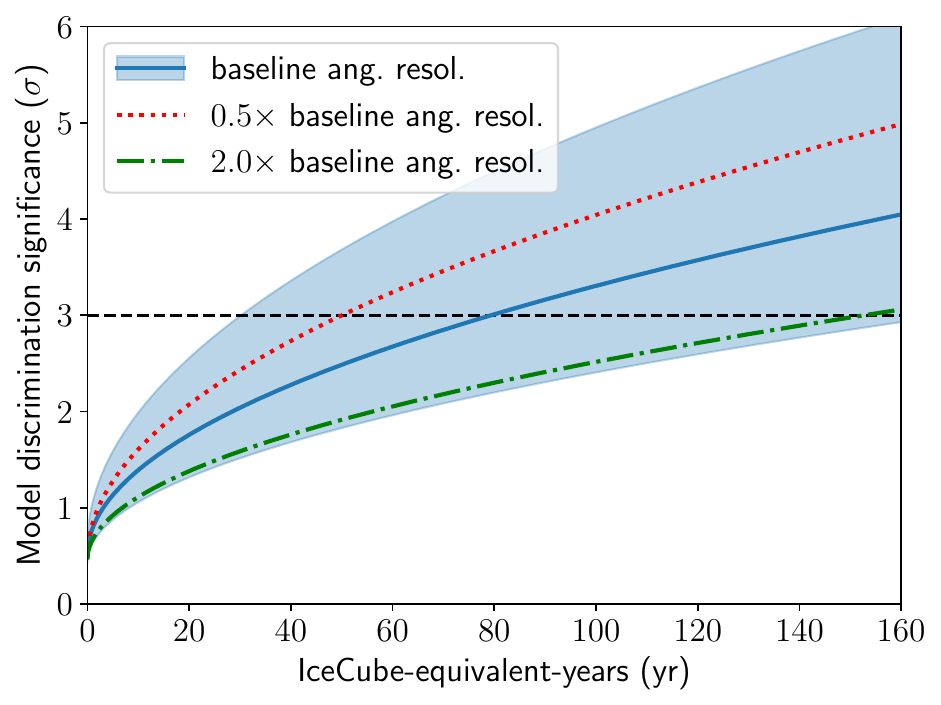}
\caption{Discrimination significance between gas distribution models built from the CS and dust mass tracers.
The dotted, solid, and dash-dotted lines correspond to different assumptions on the angular resolution. The blue line and band correspond to calculations assuming the best-fit HESS flux or its 1$\sigma$ uncertainty envelope as input.}
\label{fig:significance}
\end{figure}

The results we obtain are shown in \figurename~\ref{fig:significance}. The model discrimination significance increases approximately with the square root of the exposure, in line with the expected scaling of statistical uncertainties. 
Our baseline result reaches a $3\sigma$ discrimination power for the best-fit HESS energy spectrum after about 80 IceCube-equivalent-years.
This is an exposure we expect to accumulate within the next three decades.

Along with the exposure, the other key experimental parameter for this analysis is the angular resolution. Our baseline result is based on the projected energy-dependent angular resolution of KM3NeT \cite{KM3NeT2024EPJC}. However, to illustrate the impact of this parameter as well as to motivate experimental efforts to further improve it, we included 
the sensitivity projections for a resolution increased or reduced by a factor of 2 in \figurename~\ref{fig:significance}. Changes of this magnitude substantially affect our sensitivity projections, increasing or reducing the significance by about 1$\sigma$.

To quantify how our results are affected by assumptions on the CMZ neutrino spectrum, we have repeated our calculations for several scenarios. Our baseline calculations rely on the neutrino spectra following a pure power law aligned with the gamma-ray spectrum. A first source of uncertainty is related to the extrapolation of the HESS best-fit result, whose uncertainty is displayed with the blue band in \figurename~\ref{fig:significance}. Additional sources of uncertainties are related to the extrapolation model, which instead of a pure power law might have an energy cut-off.
To assess this second source of uncertainties, we have repeated our calculations for a scan of cut-off values. Our selected values effectively cover a variety of models. For instance, the KRA$\gamma$-50 and KRA$\gamma$-5 models are well approximated by a power law with an exponential cut-off of 1.6\,PeV and 360\,TeV \citep{Gaggero15PRD}. 
Similarly, we also considered a simultaneous fit of all data from HESS, MAGIC, and HAWC, which provides 80\,TeV as best estimate for the cut-off value (see \figurename~\ref{fig:gamma_ray_obs}).
Our calculations show that cut-off values of 1.6\,PeV, 360\,TeV, and 80\,TeV increase the exposure required to achieve a 3$\sigma$ discrimination to 85, 90, and 125 IceCube-equivalent-years, respectively.
We find that an exponential energy cut-off larger than 1\,PeV does not affect our sensitivity projections significantly, as neutrino events above this energy are rare.

In conclusion, our work shows how neutrino observations will provide valuable insights into the gas distribution of the CMZ and the fraction of gamma-rays generated through leptonic and hadronic processes in the Galactic Centre. 
While achieving high-significance model discrimination may require two decades or more depending on the assumptions, neutrino telescopes will ultimately provide a fully independent and solid test of the gas density inferred from mass tracers. This improved measurement of gas properties will facilitate the development of star-formation models in the CMZ, and this knowledge will be generalised to model galaxies beyond the Milky Way.

In addition, the next decade will witness a new wave of neutrino telescopes coming online alongside the Cherenkov Telescope Array Observatory (CTAO). CTAO will enable a step-change in the precision of gamma-ray measurements of the CMZ~\cite{2018}. In particular, by detecting photons up to 300\,TeV, CTAO will provide crucial insights into the presence of a possible cut-off affecting the neutrino spectra above the range explored by HESS and HAWC. At the same time, the next generation of neutrino telescopes will provide constraints on the gamma-ray flux due to hadronic processes, potentially leading to the identification of gamma-ray sources powered by leptonic processes. The reciprocal feedback between these two messengers will drive transformative progress and push the frontiers of multimessenger astronomy. 

\textit{Acknowledgements}---The authors would like to thank Foteini Oikonomou, Kate Pattle, and Sheng-Jun Lin for useful discussions.
We thank the referees for their valuable feedback.
We thank ESA for making \emph{Herschel} data available, and NAOJ for making NRO data available.
We are also grateful to the PLE$\nu$M group 
(\url{http://github.com/PLEnuM-group/Plenum})
for making their software open source.
This work has been supported by the Cosmoparticle Initiative of UCL. PCWL is supported by a UCL Graduate Research Scholarship and a UCL Overseas Research Scholarship. MA acknowledges support from the Science and Technology Facilities Council, part of the UK Research and Innovation (Grant No. ST/T004169/1). ERO is an international research fellow under the Postdoctoral Fellowship of the Japan Society for the Promotion of Science (JSPS), supported by JSPS KAKENHI Grant Number JP22F22327, and also acknowledges support from the RIKEN Special Postdoctoral Researcher Program for junior scientists. 

\bibliographystyle{h-physrev}
\bibliography{neutrino}

\end{document}